\documentclass[lettersize,journal]{IEEEtran}
\usepackage{amsmath,amsfonts}
\usepackage{algorithmic}
\usepackage{algorithm}
\usepackage{array}
\usepackage[caption=false,font=normalsize,labelfont=sf,textfont=sf]{subfig}
\usepackage{textcomp}
\usepackage{stfloats}
\usepackage{url}
\usepackage{verbatim}
\usepackage{graphicx}
\usepackage{cite}
\begin{document}

\title{3D Object Positioning Using Differentiable Multimodal Learning}

\author{{Sean Zanyk-McLean,~\IEEEmembership{~The University of Texas at Austin}} \and \\ {Krishna Kumar,~\IEEEmembership{~The University of Texas at Austin}} \and \\ {Paul Navrátil,~\IEEEmembership{~The University of Texas at Austin}} }



\maketitle

\begin{abstract}
This article describes a multi-modal method using simulated Lidar data via ray tracing and image pixel loss with differentiable rendering to optimize an object's position with respect to an observer or some referential objects in a computer graphics scene. Object position optimization is completed using gradient descent with the loss function being influenced by both modalities. Typical object placement optimization is done using image pixel loss with differentiable rendering only, this work shows the use of a second modality (Lidar) leads to faster convergence. This method of fusing sensor input presents a potential usefulness for autonomous vehicles, as these methods can be used to establish the locations of multiple actors in a scene. This article also presents a method for the simulation of multiple types of data to be used in the training of autonomous vehicles.
\end{abstract}

\begin{IEEEkeywords}
Differentiable Rendering, Inverse Rendering, Lidar, Object Position Optimization, Gradient Descent, Sensor Fusion
\end{IEEEkeywords}

\section{Introduction}
Differentiable rendering is an emerging technique in computer graphics that enables the calculation of gradients with respect to parameters in a 3D rendering pipeline. Recent advances in physics-based differentiable rendering allow researchers to generate realistic images by accurately representing light propagation through a scene. Differentiable rendering enables solving complex inverse rendering problems such as the optimization of 3D scene parameters via gradient descent~\cite{DBLP:journals/corr/abs-2006-12057}. 

Simultaneously, the integration of data from multiple sensor modalities, termed multi-modal sensor fusion, has captured researchers' attention due to its significance in the development of autonomous vehicles~\cite{DBLP:journals/corr/abs-2202-02703}.  For instance, researchers fuse vision systems and Lidar (light detection and ranging) to enhance autonomous driving capabilities. Lidar is a remote sensing technology that uses laser pulses to measure distances to objects and create precise 3D representations of the surrounding environment. It works by emitting a laser beam that bounces off objects and returns to the sensor, allowing the creation of high-resolution 3D maps of the environment. 

Lidar technology can be used with 3D object detection algorithms to identify and classify objects in the environment. 3D object detection algorithms analyze the Lidar data to identify the location, size, and shape of objects in the environment, such as cars or pedestrians. Lidar data can be combined with other sensor data, such as vision systems and radar, to provide a more complete and accurate picture of the environment and to help identify and track objects in real time. This technology has numerous applications, including self-driving cars, robotics, and urban planning.

We introduce a method for optimizing the placement of objects in a 3D graphics scene relative to a specified viewpoint leveraging multi-modal sensor fusion and differentiable rendering. In this inverse rendering setup, an object starts in an initial position within a 3D graphics scene. The goal is to transform the object's position to a predefined target location through optimization. The optimization hinges on an image loss comparison between the current object's rendered position (in terms of pixel values) and the target position's rendered image.

We employ differentiable rendering, which moves the object from its starting point to its target by using gradients of the rendered image relative to scene parameters. Our approach augments conventional differentiable rendering by incorporating Lidar data, allowing for 3D object detection and distance measurements from the sensor to the object. This depth sensing enhances the optimization by conveying object distances and positions relative to the viewpoint. The optimized object can be any visible element in the scene, such as a car or light. Alternatively, we can optimize for the observer location (camera) while visible objects remain fixed. The multi-modal fusion of vision and Lidar sensing facilitates precise 3D object positioning to match desired viewpoints.

\section{Related Works}
Our multimodal differentiable rendering method builds upon prior work on inverse graphics and sensor fusion. Mitsuba \cite{jakob2022mitsuba3} and PyTorch3D \cite{ravi2020pytorch3d} support gradient-based optimization of inverse rendering with ray tracing. These tools are enhanced and made possible by many advances in gradient-based methods for rendering, including Differentiable Monte Carlo Ray Tracing through Edge Sampling \cite{Li:2018:DMC}.

Differentiable rendering has been used in many ways, for example in Zakharov et al. ~\cite{zakharov2020autolabeling} differentiable rendering was used to predict shapes and poses from image patches. Their pipeline combines vision and geometry losses for multimodal optimization.

There is a rich literature on sensor fusion for 3D understanding. For instance, Perception-Aware Multi-Sensor Fusion for 3D Lidar Semantic Segmentation \cite{zhuang2021perceptionaware} fuses appearance information from RGB images and spatial-depth information from point clouds to improve semantics segmentation.

Our method integrates insights from prior work to address multimodal inverse rendering tasks. The flexibility of differentiable rendering enables joint optimization over multiple data sources and loss functions. 

\section{Methods}

At the core of our optimization is Mitsuba, a research-oriented differentiable renderer. Mitsuba leverages automatic differentiation via Dr.Jit~\cite{Jakob2022DrJit} to enable gradient-based inverse rendering. We use Mitsuba to simulate both RGB images and Lidar data from 3D scenes. The built-in optimizers allow joint training over multimodal losses. To render a scene in Mitsuba, users specify parameters including integrator, max depth, samples per pixel, and sampler type. We use path tracing with a bidirectional path tracer integrator. The number of samples per pixel is set to 16 to reduce Monte Carlo noise. The gradients from the renderer are used to iteratively refine scene parameters like camera pose, lighting, and materials using the Adam optimizer. 

We use Mitsuba's Reparameterized Path Replay Backpropagation integrator~\cite{10.1145/3450626.3459804} for differentiable rendering. This technique performs integration over high-dimensional lighting and material spaces. It provides an efficient path tracer that handles discontinuities in 3D scenes. We set the max path depth to 3 for our experiments and used 16 samples per pixel. These values balance computation time and rendering quality for our purposes. Using more samples and higher max depth yields more photorealistic results at increased cost. We set the sampler type to be independent of uncorrelated noise.  Experiment resources and code can be found on Github.\footnote{\url{https://github.com/szanykmclean/differentiable_multimodal_learning}}. 

We use a simple scene containing a car model on a homogeneous background for our experiments. This setup is motivated by autonomous driving applications, where 3D detectors are often trained to locate cars. The objective is to optimize the camera pose with respect to the stationary car object. We fix the car's position and orientation and optimize the camera location and orientation to match target renderings. This inverse rendering approach could also be applied to optimize object poses given fixed camera intrinsic and extrinsic parameters. While simple, this scene captures key challenges in multimodal inverse rendering. Optimizing camera pose requires reasoning about viewpoint, visibility, lighting, and materials. The homogeneous background enables the isolation of the car as the primary focus. More complex scenes could incorporate detailed environments and multiple objects. The differentiable rendering approach provides a principled methodology to handle complex scenarios with multiple objects, occlusion, and background clutter. Overall, this controlled setup provides a strong testbed for multifaceted inverse rendering of a central 3D object.

\subsection{Lidar Data}
In order to generate Lidar data, we use the built-in ray-tracing functionality of the Reparameterized Path Replay Backpropagation Integrator. During rendering, the ray intersections at a depth of 0 are recorded and written to a text file. Each ray intersection is an instance of light bouncing off an object in the scene, and an $(x, y, z)$ coordinate is recorded. All the intersection points are used together to create a simple point cloud without intensity data. This point cloud effectively simulates Lidar data, a common data modality in autonomous driving. Lidar data allows for distance estimation. This data is used with a large trained 3D object detection network and will allow the system to utilize distance measures from the camera to the car. 

\begin{figure}[!htb]
\centering
\includegraphics[width=3.0in]{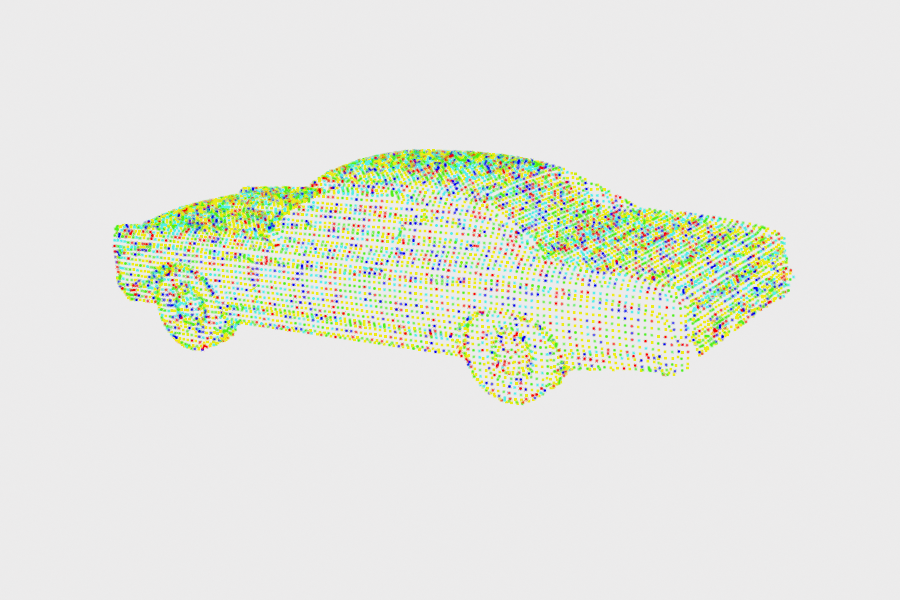}
\caption{Lidar data generated via Mitsuba ray-tracing. The variation in the colors of the points is used to improve the visualization of the Lidar data.}
\label{fig_1}
\end{figure}

\subsection{3D Object Detection}
After the Lidar data is generated, a 3D object detection algorithm is used to generate a bounding box for the car located in the scene. The detection algorithm used is PointPillars \cite{DBLP:journals/corr/abs-1812-05784} which is an encoder that utilizes PointNets to learn a representation of point clouds organized in vertical columns (pillars). The algorithm is pre-trained on the KITTI 3D object detection dataset \cite{Geiger2012CVPR}, a large and commonly used autonomous vehicle dataset made available by the Intelligent Systems Lab Organization \cite{Zhou2018}. This pre-training allows the algorithm to detect objects during inference, specifically cars, buses, and pedestrians. Point Pillars inference is applied to the text file containing Lidar data in $(x, y, z)$ format generated via the previous section. 

The algorithm detects a car object in the scene and is able to generate a bounding box around the object. This algorithm may have slight variations depending on the target camera location and a resulting point cloud of Lidar data. It will not be effective during every inference at detecting the car. This makes intuitive sense as car location being far away or at unique locations with respect to the camera will alter the Lidar data and potentially cause it to become unrecognizable to the system. In practice, the system works well in simple scenes, and the method presented will utilize the bounding box with the highest confidence score for a predicted car class. This establishes the $(x, y, z)$ location of the car in the scene and the distance from the camera location to the car.

\begin{figure}[!htb]
\centering
\includegraphics[width=3.0in]{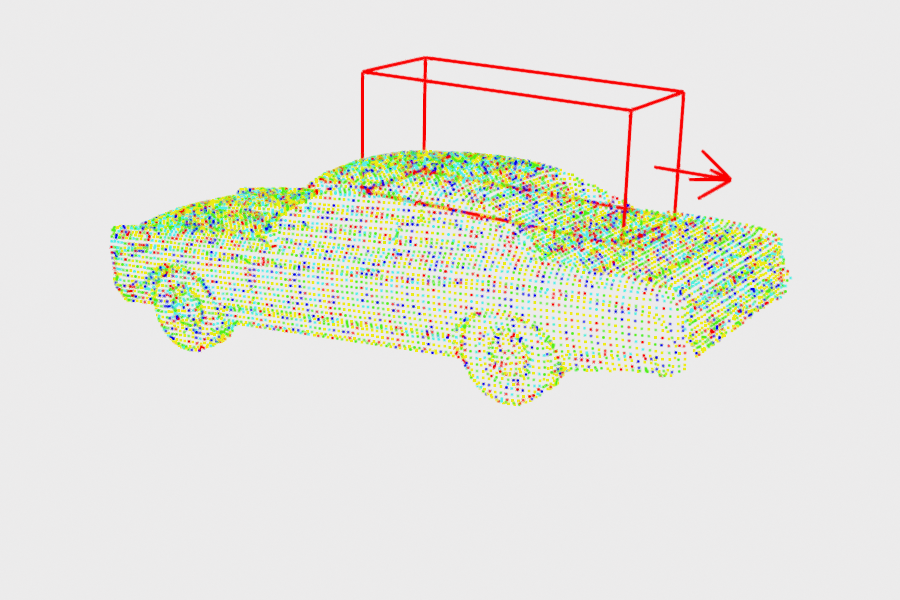}
\caption{PointPillars object detection using Lidar data. The bounding box and arrow show a detected car object oriented towards the direction of the arrow.}
\label{fig_2}
\end{figure}

\subsection{Initial and Target Camera Locations}
The experiment's goal is position optimization, which involves moving an object from an initial position to a target position with respect to the objects and observers in the scene. An initial and target object (camera) location was used with the simple 3D computer graphics scene containing one car object to experiment and show the utility of using multi-modal data for object position optimization. The initial camera location has an $(x, y, z)$ coordinate of $(20, 13, 23)$ and the target camera location is $(8, 5, 14)$. The target camera location was translated by $(12, 8, 9)$ to be used at the start of the optimization loop. The values are unitless as this is a simulated computer vision scene. In the images below, one can see that the initial camera location is much further away from the car and that the car is not centered in the camera view. The translation of the camera location from the target to the initial location at the start of the optimization moves the camera further away from the car in the scene. The target camera location is a distance of 16.93 units away from the center of the established bounding box for the car. The initial location is further away, with a distance of 32.51 units from the car. This is almost twice the distance from the target location to the car. This initial location was chosen to effectively show how the distance loss with Lidar data can improve the optimization. No rotations were applied to the target camera location in order to keep the optimization simple and allow for convergence.

\begin{figure}[!htb]
\centering
\includegraphics[width=3.0in]{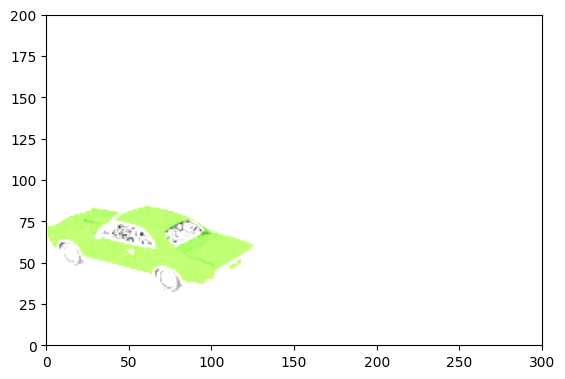}
\vspace{.35cm}
\includegraphics[width=3.0in]{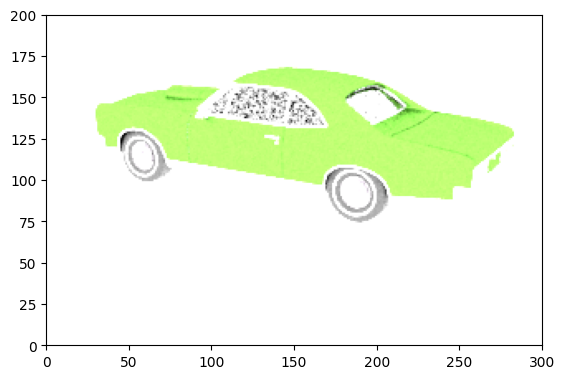}

\caption{Initial camera location is shown in the top image. The target camera location is shown in the bottom image.}
\label{fig_3}
\end{figure}

\subsection{Object Position Optimization}
In order to move the object position to the target position, the system utilizes Adam \cite{kingma2014adam}, a first-order gradient-based optimization method with a learning rate of 0.15. This learning rate was selected as it resulted in optimal convergence given the desired iteration limit. For the experiments, 30 iterations of gradient descent were used, and at each step, the object's position was moved based on the gradients of the loss function. In each iteration, the computed transformation based on the previous gradients is first applied to the object's position. A new image is rendered based on this new object's position using Mitsuba. Then, the new $(x, y, z)$ object's position is compared to the 3D object's detected car-bounding box location. The distance between the current object's position and the car is computed. Then, the current object's distance from the car is compared to the target object's distance from the car. A loss function is used to compare these two values, and this loss is combined with image pixel-wise loss between the target ending image and the current object's location rendered image. These two losses are combined to help steer the object location to the target through each step of gradient descent and help weigh the optimization to consider not only the gradients of the pixels with respect to scene parameters but also the current distance of the object from the car using the simulated Lidar data. This aims to help the object avoid moving in the wrong direction during the optimization as the camera can often lose the car off-screen when searching for a lower image loss.

\subsection{Joint Loss Function}
One component of the joint loss function is distance loss. Distance loss is derived from the distance formula for Euclidean distance. Two distances are calculated and compared in value. The first distance calculated is from the current object location which is denoted as $(x_{c}, y_{c}, z_{c})$ to the center of the bounding box which is denoted as $(x_{b}, y_{b}, z_{b})$ and was detected via the previous object detection stage. This distance is denoted as $d_{c}$. The second distance calculated is from the target object location which is denoted as $(x_{t}, y_{t}, z_{t})$ to the bounding box is also computed using the same formula and is denoted as $d_{t}$: $$d_{c} = \sqrt {\left( {x_{c} - x_{b} } \right)^2 + \left( {y_{c} - y_{b} } \right)^2 + \left( {z_{c} - z_{b} } \right)^2 }$$
$$d_{t} = \sqrt {\left( {x_{t} - x_{b} } \right)^2 + \left( {y_{t} - y_{b} } \right)^2 + \left( {z_{t} - z_{b} } \right)^2 }$$\\

These two distances are compared using Root Mean Squared Error (RMSE) with a scalar value $\alpha$ to calculate the loss, $L_{d}$. The other component of the joint loss function is image loss. The number of pixels in an image rendered during the optimization is denoted as $N$ and is simply defined as $N=l\cdot{w}$ where $l$ is the length of the image and $w$ is the width. In the experiments, $l=200$ and $w=300$. At each step during the optimization, image loss is computed by comparing the currently rendered image to the target image at each corresponding pixel value. The image loss function outputs one scalar value which is the Mean Squared Error (MSE). In the loss function, $x_j$ is the current image pixel value at index $j$ and $\hat{x}_j$ is the target image pixel value at index $j$. The resulting function for image loss is defined as $L_{i}$. Both loss components are defined below:

$$L_{d}=\alpha\cdot\sqrt{(d_c-d_t)^2}$$
$$L_{i}=\frac{\sum_{j=0}^{N}(x_j-\hat{x}_j)^2}{N}$$

The scalar $\alpha$ is used to help weigh the importance of distance during the optimization. Finally, the joint loss function is defined as:
$$ L = L_{i} + L_{d}$$

This loss function will be used in multiple experiments to assess the usefulness of the multi-modal method and will be compared with baseline optimization methods. One experiment will utilize this joint loss $L$ during the entire optimization. Another experiment which is shown in Fig. 4 will utilize a two-stage loss function that utilizes the joint loss function during optimization while the object (camera) is a user-defined threshold away from the target distance, then the optimization will switch to the second stage where the simple image loss will be used to optimize the location of the car in the image.

\begin{figure}[!htb]
\centering
\includegraphics[width=3.3in]{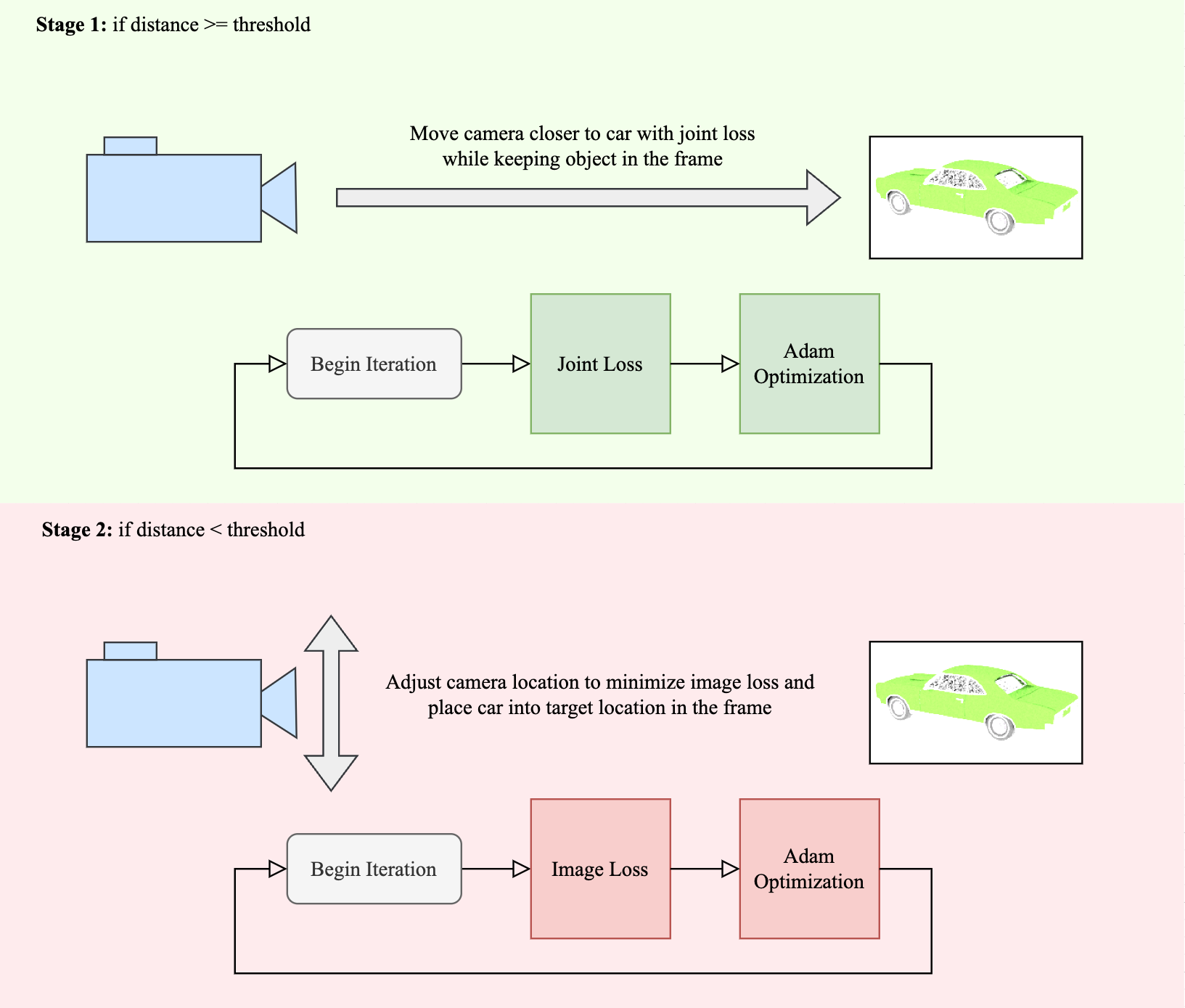}
\caption{Two-stage joint loss diagram.}
\label{fig_4}
\end{figure}

\section{Results}
Experiments were conducted using the previously mentioned methods as well as baseline methods for comparison and establishing performance. After running four separate experiments with different loss functions, the two-stage loss method is able to converge in the shortest amount of iterations and to the best location. This experiment is described below and utilizes the presented method of multi-modal optimization.

\subsection{Image Loss}
One experiment conducted was to use only image loss for the inverse rendering problem. This is a common out-of-the-box method and establishes a baseline performance. The results for this method clearly show that the 
optimization process will at first move further away from the target object location due to the gradients of the image with respect to the scene parameters. This is clearly sub-optimal behavior. Towards the end of the optimization, the camera location is moving in the correct direction, however, it is taking several iterations to begin converging towards the appropriate direction.

\begin{figure}[!htb]
\centering
\includegraphics[width=3.25in]{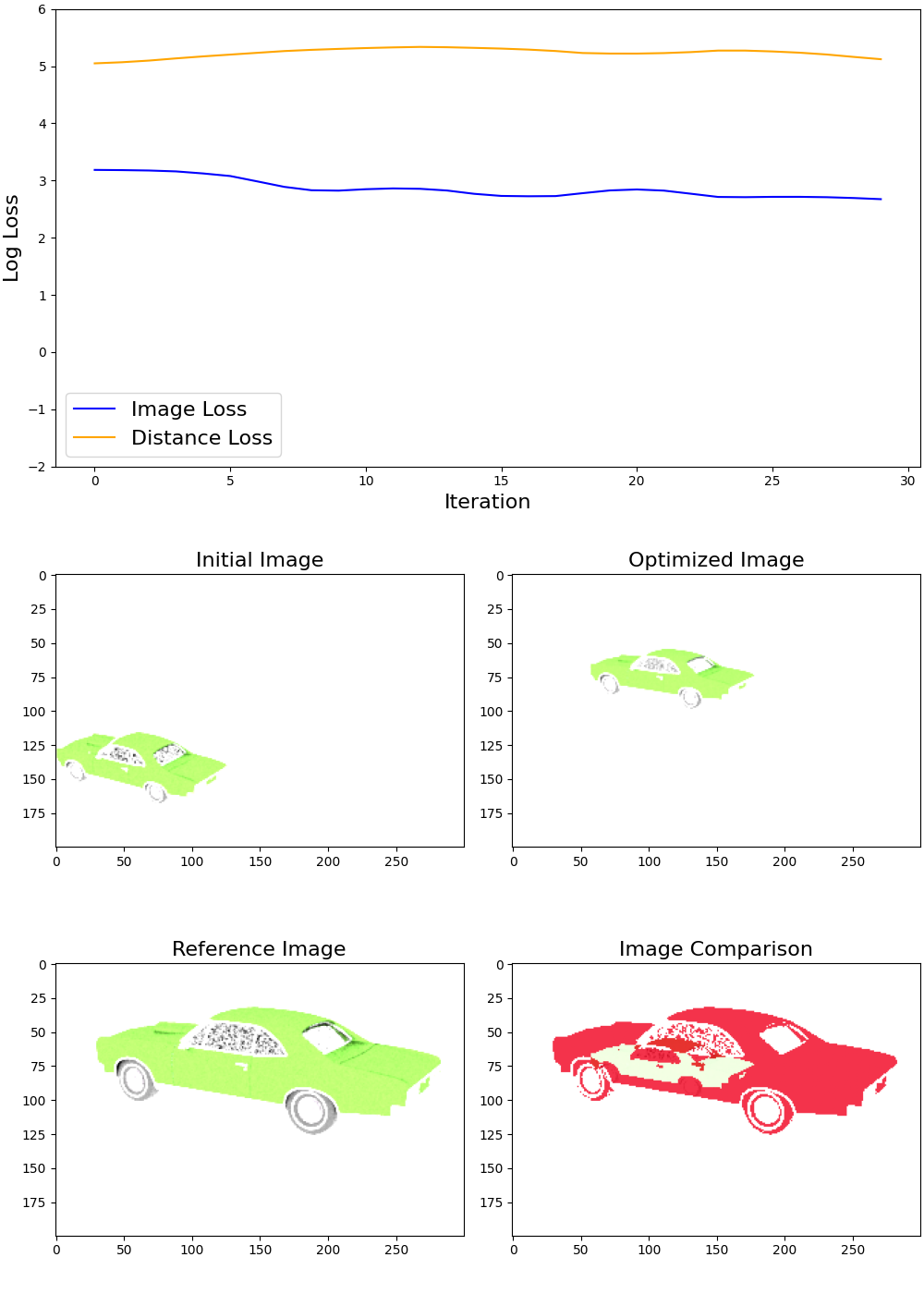}
\caption{Image loss optimization. The final distance from the car bounding box to the camera was 33.7 compared to a target distance of 16.9.}
\label{fig_5}
\end{figure}

\subsection{Distance Loss}
Another experiment conducted was to use only distance loss for the inverse rendering problem. If the translation of the target image to the initial image was simply an equal scaled move in every $(x, y, z)$ direction, then, in theory, this method would work very efficiently and be optimal. However, from the results, it is clear this is sub-optimal. The optimization only uses distance as the guiding metric for object position so the location and pixels of the car are clearly ignored and the object will only move towards the target with the goal of finding the optimal distance from the car, but will not be able to find the target location.

\begin{figure}[!htb]
\centering
\includegraphics[width=3.25in]{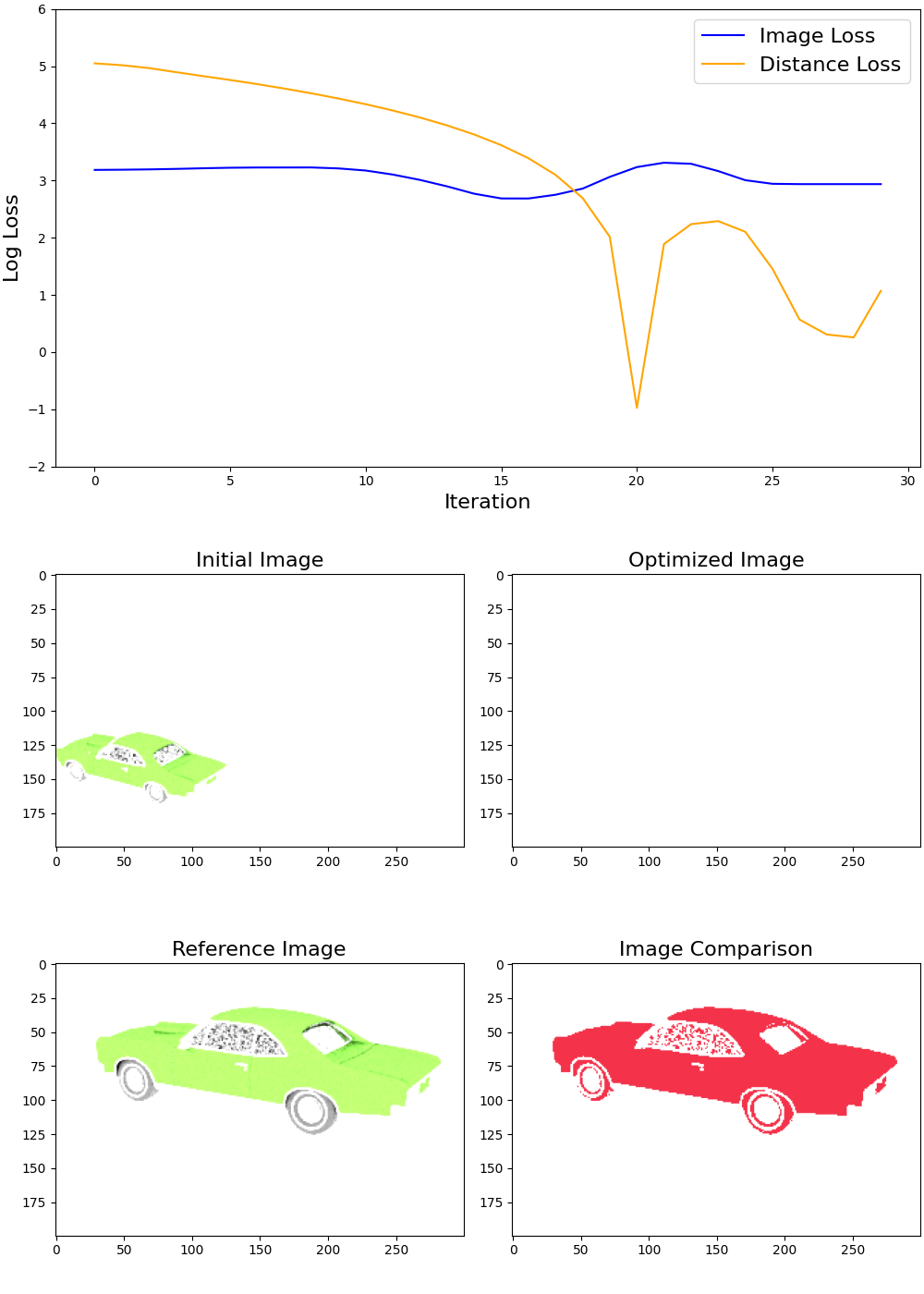}
\caption{Distance loss optimization. The final distance from the car bounding box to the camera was 17.2 compared to a target distance of 16.9.}
\label{fig_6}
\end{figure}

\subsection{Joint Loss}
To test this presented method of camera optimization, an experiment using the joint loss method $L$ which takes both image loss and distance loss into account is presented here. Clearly, the use of both image loss and distance loss to optimize object position is leading to lower image loss and more optimal camera location than the previous two methods. It is also leading to faster optimization. One issue that is clear with this method is that distance loss as a guide for optimization works well when the camera is relatively far away from the target distance. However, when the camera is already close to the target distance, this part of the joint loss is seemingly forcing the optimization out of the correct location. Distance loss prevents the system from doing the necessary small $(x, y, z)$ coordinate transformations that may lead to higher distance loss but allows for the image loss optimization to find the correct location. This is evident by the car in the optimized image being slightly out of the frame in Fig. 7. The value of $\alpha$ selected for this optimization is 10. This value was selected based on experimentation and was chosen to balance out the two loss components of the loss function. The image comparison heat map shows some overlap between the initial and target car locations, however, there is still room for improvement.

\begin{figure}[!htb]
\centering
\includegraphics[width=3.25in]{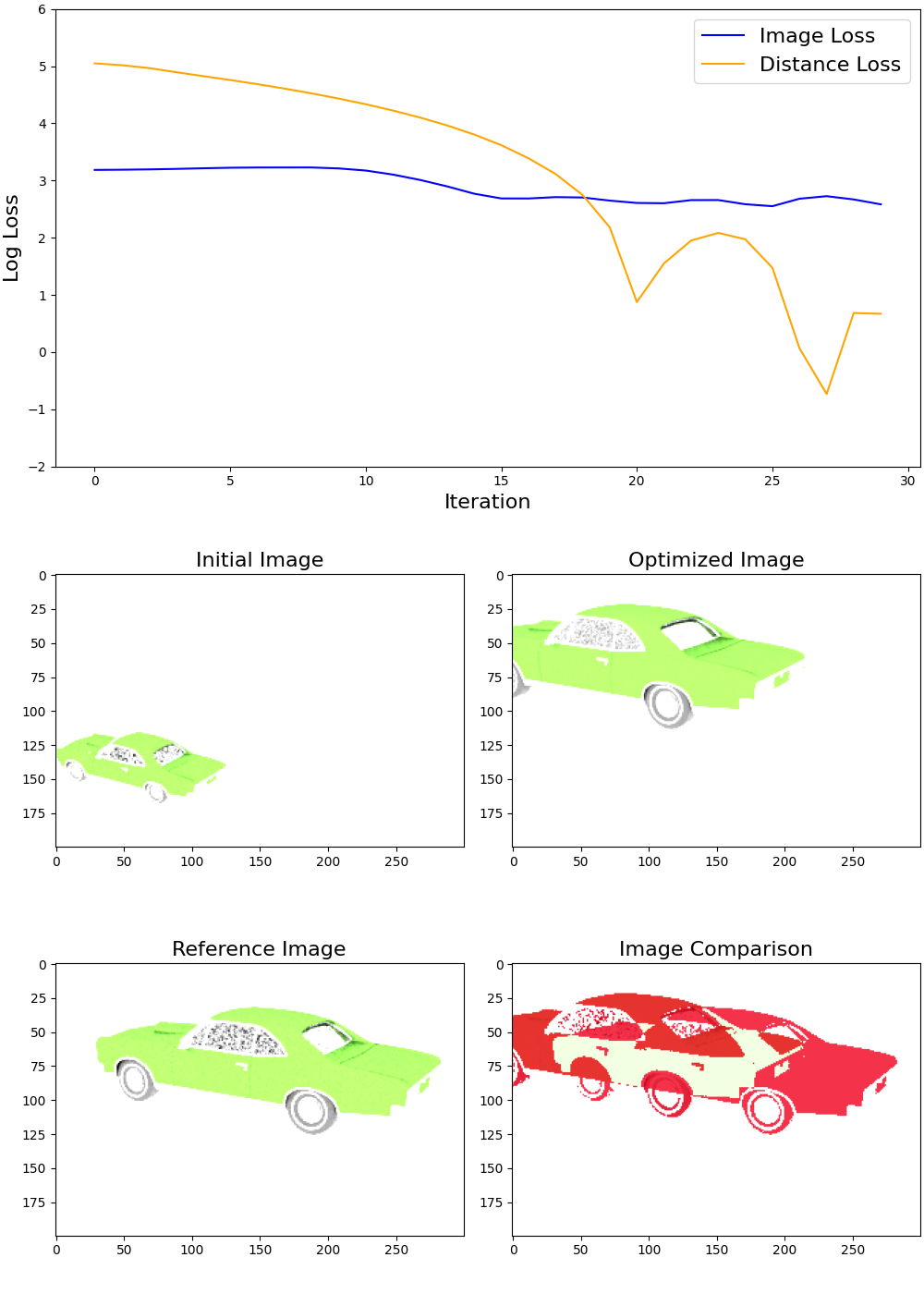}
\caption{Joint loss optimization. The final distance from the car bounding box to the camera was 17.1 compared to a target distance of 16.9.}
\label{fig_7}
\end{figure}

\subsection{Two-Stage Joint Loss}
To solve the issues with the joint loss optimization, a new experiment was conducted using the two-stage loss. For the first part of the optimization, the joint loss method was used. Once the camera location reaches a distance that is less than a user-defined distance threshold, the optimization will set $\alpha$ to 0 and will then be guided only by the image loss method from the first experiment. Before the threshold is reached, the value of $\alpha$ is again set to 10. The threshold distance where $\alpha$ is set to 0 is also user selected and for this experiment, it was set to 2.0 units. This threshold was selected after experimenting with thresholds in the range of 1.0 to 5.0 and discovering that this value led to optimal performance. Setting the threshold too high led to slower convergence and setting it too low led to a similar performance as the Joint Loss method. This method avoids the issues of using distance loss from a close camera distance and allows for fast optimization with very optimal image loss and therefore camera location. It is clear from the results that this method is the most effective for differential camera optimization. The image comparison heat map shown in Fig. 8 helps show the optimal performance of this method as the cars nearly perfectly overlap and achieve the best performance of any of the experiments.

\begin{figure}[!h]
\centering
\includegraphics[width=3.25in]{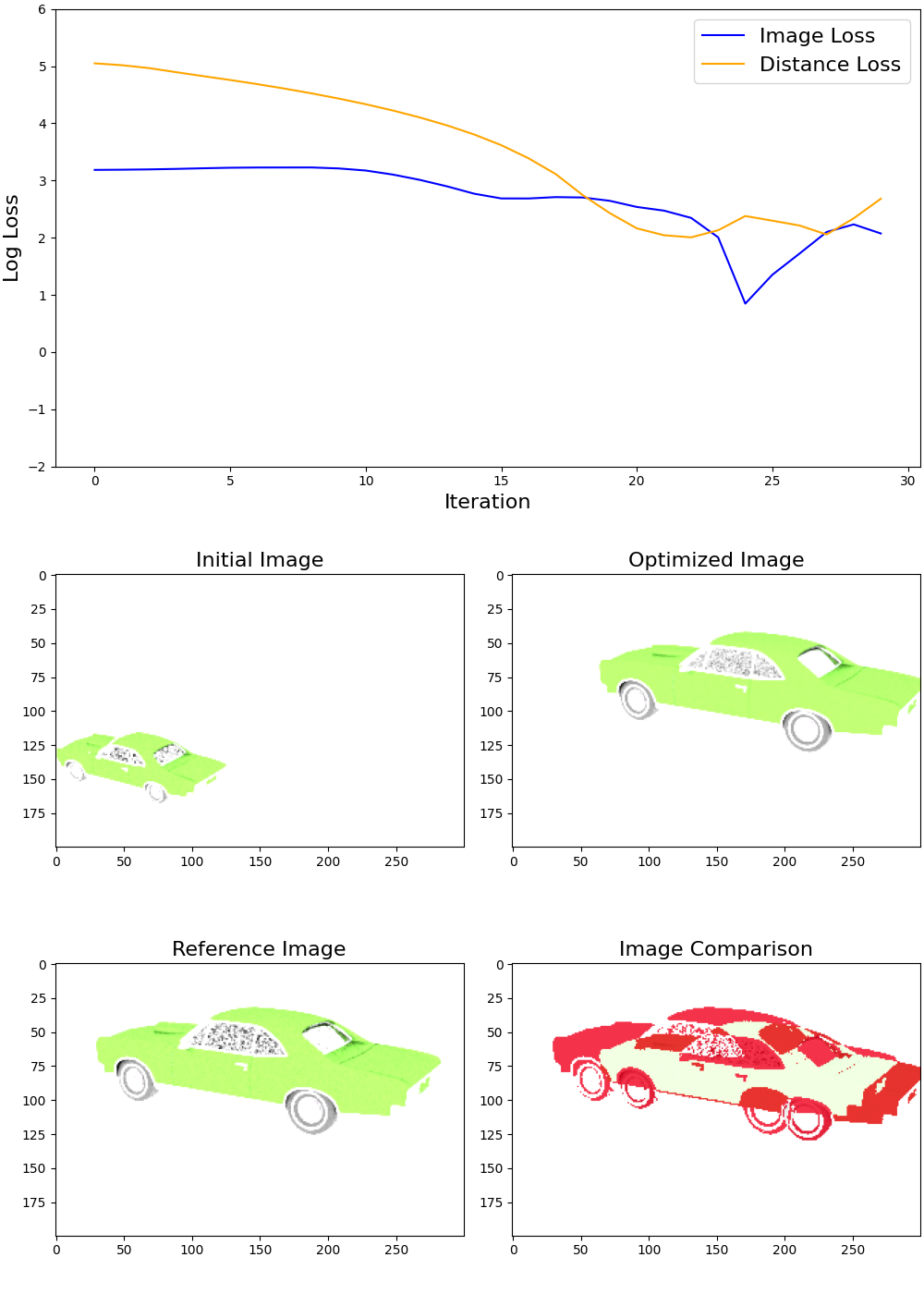}
\caption{Two-stage joint loss optimization. The final distance from the car bounding box to the camera was 18.4 compared to a target distance of 16.9.}
\label{fig_8}
\end{figure}

\section{Discussion}
The results in the previous section clearly show there are advantages to using a joint loss function that not only considers image pixel values but also considers an Euclidean distance metric. The results obtained are heavily variant based on scene selection, parameters, and object detection quality. For instance, while testing this method with more realistic images and scenes, the object position optimization would often fail to converge at all. This is an issue with non-uniform backgrounds and images that make it extremely hard for optimal solutions to be found. This is suspected to be because of comparison of pixel values in non-homogeneous backgrounds can lead to image pixel losses that are heavily variant and can significantly increase loss values even when the object position optimization is moving towards the correct location. In other words, the complicated and realistic scenes cause the data to be very noisy and lack a sufficient signal for optimization. 

\begin{table}[!htb]
    \centering
    \caption{Comparison of Results After Optimization.}
    \begin{tabular}{l|l l l}
          & Loss & Camera & Distance To \\
         Method & (Img / Dist) & Position $(x, y, z)$ & (Car / Target) \\\hline 
         Image & 14.5 / 167 & $(18.5, 9.35, 27.5)$ & 33.7 / 17.6 \\ 
         Distance & 18.9 / 2.92 & $(10.0, -0.15, 13.0)$ & 17.2 / 5.61 \\
         Joint & 13.3 / \textbf{1.95} & $(10.0, 4.65, 13.0)$ & 17.1 / 2.26\\
         Two-Stage & \textbf{7.96} / 14.6 & $(7.60, 5.55, 15.4)$ & 18.4 / 1.56 \\ \hline
         Target &  & $(8.0, 5.0, 14.0)$ & 16.9 \\\hline 
    \end{tabular}
\label{tab:default}
\label{fig_2}
\end{table}

\subsection{Scene Selection}
The scene used in the experiments and optimization used a translation to convert the target object position to the initial object position, however, no rotation was applied to the initial scene. This was purposefully left out to avoid too much complexity in the object optimization. Adding rotation in as a parameter to optimize for the camera can cause the optimization to fail where it otherwise might have successfully converged. Optimization of many scene parameters at once leads to difficulties and this is an area that could be explored further. 

One important finding during experimentation is that object position optimization using only image data and even using a multi-modal method is more difficult on scenes with non-homogeneous backgrounds as well as scenes that do not have sufficient lighting. These issues seem to be derived from using MSE of pixel data which will not be able to establish useful loss values if most of the pixels appear relatively dark or non-homogeneous and have similar values. For instance, the same car scene which is given a much more realistic look using a background object and lighting led to difficulties converging during experimentation. The object position optimization will often lose the referential object entirely. This presents the importance of completing image processing and background masking with segmentation before object position optimization. In order to use this method in realistic scenes, pre-processing of images is a potential area that can be explored further.

\subsection{Hyperparameter Selection}
Results are also heavily affected by the hyperparameters. User-selected hyperparameters for these experiments include setting a learning rate of 0.15, setting a sample per pixel value of 16, choosing alpha to be 10, and setting a threshold to 2.0 for the two-stage loss. The selection of these hyperparameters was tuned by running many experiments and establishing baseline performance results. One further focus that could be implemented related to this method is establishing rules and metrics for how to choose alpha values effectively as well as threshold values in the two-stage method. Finding the right proportion of loss between the image and distance is important for finding an optimal convergence. If one is overweighted, then the solution with converge to a solution similar to either the image loss only or the distance loss only. One difficulty with this selection of alpha is that image loss values can vary heavily from scene to scene.

\subsection{3D Object Detection}
PointPillars was used for object detection and bounding box generation, which is a very important part of this system. Establishing accurate bounding boxes on the simulated Lidar data is important because a lack of accuracy of the car object in the scene will lead to inaccurate distance measurements and will lead to convergence at a potentially incorrect location. For purposes of the experiment, both the initial and target camera locations used a center bounding box location generated from the target camera location. This was done to avoid issues of incorrect 3D object detection when changing location to the initial location. The PointPillars algorithm is heavily dependent on the point cloud data it is given. Changing location and generating point clouds from the new location can cause the algorithm to miss objects, such as the car in the experiments. 

The system could also work by using 3D object detection and bounding box establishment from both locations, it is however subject to more noise and potential differences in location or objects detected. Furthermore, the PointPillars algorithm was pre-trained on the KITTI dataset and it is clear from Fig. 2 that the bounding box is not perfectly enclosing the car object. This is most likely due to differences in the training data and the Lidar data given at inference time. In order to further optimize this portion of the system, more robust 3D object detection algorithms could be tested and used. Another issue with these algorithms is they can detect multiple instances of the same object where there is only one object and this was seen during testing. In order to offset this, the reference object was established as the object detected with the highest confidence and the correct classification.

\section{Conclusion}
This paper presents a novel method for performing differentiable multi-modal object position optimization. The method utilizes both image data and synthesized Lidar data to inform the gradients during the optimization and leads to better convergence to the target object position in the experiments when compared with baseline methods. This method furthers the performance of inverse rendering techniques and displays ways to fuse multiple modalities to improve performance. Applications of this technology could include autonomous driving systems and robotics. These methods could improve state estimation and scene understanding for multiple vehicles in proximity to each other, especially if optimization can be completed in a fast and computationally efficient manner on embedded devices. 

\nocite{*}
\bibliographystyle{IEEEtran}
\bibliography{paper}

\end{document}